\documentclass[11pt]{article}

\usepackage{graphicx,amsfonts,amsbsy, amssymb, amsthm, amsmath, mathrsfs}

\newcommand{\rmd}{{\mathrm d}}
\newcommand{\rme}{{\mathrm e}}
\newcommand{\rmi}{{\mathrm i}}

\DeclareSymbolFont{lettersA}{U}{pxmia}{m}{it}

\DeclareMathSymbol{\alphaup}{\mathord}{lettersA}{"0B}
\DeclareMathSymbol{\betaup}{\mathord}{lettersA}{"0C}
\DeclareMathSymbol{\gammaup}{\mathord}{lettersA}{"0D}
\DeclareMathSymbol{\deltaup}{\mathord}{lettersA}{"0E}
\DeclareMathSymbol{\epsilonup}{\mathord}{lettersA}{"22}
\DeclareMathSymbol{\zetaup}{\mathord}{lettersA}{"10}
\DeclareMathSymbol{\etaup}{\mathord}{lettersA}{"11}
\DeclareMathSymbol{\thetaup}{\mathord}{lettersA}{"12}
\DeclareMathSymbol{\iotaup}{\mathord}{lettersA}{"13}
\DeclareMathSymbol{\kappaup}{\mathord}{lettersA}{"14}
\DeclareMathSymbol{\lambdaup}{\mathord}{lettersA}{"15}
\DeclareMathSymbol{\muup}{\mathord}{lettersA}{"16}
\DeclareMathSymbol{\nuup}{\mathord}{lettersA}{"17}
\DeclareMathSymbol{\xiup}{\mathord}{lettersA}{"18}
\DeclareMathSymbol{\piup}{\mathord}{lettersA}{"19}
\DeclareMathSymbol{\rhoup}{\mathord}{lettersA}{"1A}
\DeclareMathSymbol{\sigmaup}{\mathord}{lettersA}{"1B}
\DeclareMathSymbol{\tauup}{\mathord}{lettersA}{"1C}
\DeclareMathSymbol{\upsilonup}{\mathord}{lettersA}{"1D}
\DeclareMathSymbol{\phiup}{\mathord}{lettersA}{"1E}
\DeclareMathSymbol{\chiup}{\mathord}{lettersA}{"1F}
\DeclareMathSymbol{\psiup}{\mathord}{lettersA}{"20}
\DeclareMathSymbol{\omegaup}{\mathord}{lettersA}{"21}

\renewcommand{\Psi}{\varPsi}
\renewcommand{\Lambda}{\varLambda}
\renewcommand{\Sigma}{\varSigma}
\renewcommand{\Gamma}{\varGamma}
\renewcommand{\Theta}{\varTheta}
\renewcommand{\Xi}{\varXi}
\renewcommand{\Pi}{\varPi}
\renewcommand{\Upsilon}{\varUpsilon}
\renewcommand{\Phi}{\varPhi}
\renewcommand{\Omega}{\varOmega}

\newcommand{\Z}{{\mathbb Z}}

\newcommand{\E}{{\mathbb E}}

\newcommand{\coloneq}{\mathbin{\hbox{\raise0.08ex\hbox{\rm :}}\!\!=}}
\newcommand{\eqcolon}{\mathbin{=\!\!\hbox{\raise0.08ex\hbox{\rm :}}}}

\renewcommand{\leq}{\leqslant}
\renewcommand{\geq}{\geqslant}
\renewcommand{\epsilon}{\varepsilon}

\newcommand{\dpdot}{{\lower0.33ex\hbox{\LARGE$\cdot$}}}

\newtheorem{theorem}{Theorem}[section]
\newtheorem{proposition}[theorem]{Proposition}
\newtheorem{definition}[theorem]{Definition}
\newtheorem{corollary}[theorem]{Corollary}
\newtheorem{lemma}[theorem]{Lemma}

\newcommand{\dimostrazione}{\noindent{\sl Proof.}\phantom{X}}
\newcommand{\dimostrazionea}[1]{\noindent{\sl Proof of #1.}\phantom{X}}
\newcommand{\finire}{\hfill$\Box$}

\newcommand{\curlyL}{{\mathscr L}}

\newcommand{\et}{equi-transmitting}

\newcommand{\ensemble}{Gaussian}
\renewcommand{\E}{G}

\begin{document}
\title{Quantum graphs where back-scattering is prohibited}
\author{J.M.~Harrison${}^1$ \and U.~Smilansky${}^2$ \and B.\ Winn${}^{1}$\\
{\protect\small\em ${}^1$ Department of Mathematics,
Texas A\&{M} University, College Station,
Texas 77843-3368, USA. }\\
{\protect\small\em ${}^2$ Department of Physics of Complex Systems,
Weizmann Institute of Science, Rehovot 76100, Israel.}}
\date{$1^{\rm st}$ August, 2007}
\maketitle
\begin{abstract}
We describe a new class of scattering matrices for quantum graphs in
which back-scattering is prohibited. We discuss some properties of
quantum graphs with these scattering matrices and explain the
advantages and interest in their study. We also provide two methods
to build the vertex scattering matrices needed for their
construction.
\end{abstract}

\thispagestyle{empty}

\section{Introduction}\label{sec:int}

It has been proposed that quantised versions of metric graphs could
be used to investigate the origin of spectral correlations in
semi-classical quantum systems \cite{kot:pot}. This programme has
borne a considerable amount of fruit in the past decade. The
culmination has been the ground-breaking work \cite{gnu:uss}
providing a mechanism to understand the fidelity of spectral
correlations to random matrix theory in a large class of so-called
quantum graphs. It is now clear that quantum graphs can be
considered at the forefront of attempts to understand the universal
spectral correlations present in quantised chaotic systems
\cite{boh:ccq,cas:otc}. For a review with ample references we refer
the reader to \cite{gnu:qga}.

One interesting new development has been the discovery that the
techniques of \cite{kot:pot} can be used to develop trace formul\ae\
and associated spectral zeta functions for {\em discrete} Laplacians
on graphs \cite{smi:qcd}. This leads to the possibility of
investigating deep connections between quantum graphs and
combinatorial spectral graph theory, hitherto unexploited. In the
present article we expand on some ideas introduced in
\cite{smi:qcd}.

We consider finite, connected graphs without loops or multiple
bonds, and which have $B$ bonds and $V$ vertices. The topology of
the graph is specified in terms of the $V\times V$ connectivity
matrix $C$, whose $p,q$ element takes the value $1(0)$ if the
vertices $p$ and $q$ are connected (not connected). We denote by
$v_j$ the valency (degree) of the vertex $j$. A graph for which all
the $v_j$ are equal to $v$,  is called a $v$-regular graph. The
metric properties are provided by the bond lengths $L_b\ , \ 1\le
b\le B$.

Spectral problems on quantum graphs are often written in terms of a
$2B\times 2B$ unitary quantum evolution matrix $U$ defined on the
vector space spanned by the {\em directed} bonds \cite{smi:qcd}. $U$
itself  can be defined in terms of the $v_j\times v_j$ unitary
scattering matrices $\sigma^{(j)}$, $j=1\dots V$ associated with
each vertex of the graph.
\begin{equation}\label{eq:defnU}
U_{(pj)(mq)}(k)=\delta_{jm}\, \sigma^{(j)}_{pq} \, \rme^{\rmi k
L_{(pj) }}
\end{equation}
where $(pj)$ is a directed bond with initial vertex $p$ and terminal
vertex $j$. The spectrum under consideration is the set of values
$\{k_n\}_{n\in\Z}$ of the spectral parameter $k$ for which
\begin{equation}
\det[I_{2B}-U(k)]=0.  \label{eq:spec:det}
\end{equation}
%
%
(Here, and elsewhere $I_n$ denotes the identity matrix of dimension
$n$.) 
A vertex scattering matrix $\sigma$ may be derived from boundary
conditions of a self-adjoint operator as in \cite{kot:pot}, or
specified {\it a priori}\/ in order to provide a wider class of
examples \cite{sch:ssq, tan:usm}, which is the approach we adopt
here.

The classical analogue of a quantum graph is a Markovian process on
the directed bonds \cite{kot:pot} with matrix $M$ of transition
probabilities related to $U$ by
$$ M_{(pj)(mq)} = |U_{(pj)(mq)}|^2.$$
Thus $M$ is a doubly stochastic matrix, and hence it has dominant
eigenvalue $1$. It can be interpreted as the classical operator
which induces a discrete time Markovian evolution on the graph. The
spectral gap $\Delta$ is the gap between the leading eigenvalue $1$
and the second-largest eigenvalue of $M$, in absolute value. It is
conjectured that the rate of convergence of spectral statistics of
quantum graphs to random matrix theory is governed by the rate of
ergodicity in this Markov evolution.  More precisely, for a sequence
of graphs with $B\to\infty$ the eigenvalues of the quantum graph are
given by random matrix theory if $B\Delta\to\infty$ \cite{tan:usm}.

We propose to study a new class of vertex scattering matrices defined in
the following way.
\begin{definition} \label{def:et}
A $v\times v$ unitary matrix $\sigma$ is \et\ if $\sigma_{pp}=0$ for
all $p$, and the off-diagonal elements have equal amplitudes;
$|\sigma_{pq}|=(v-1)^{-\frac{1}{2}}$ for $p\neq q$, where $v$ is the
valency of the vertex.
\end{definition}
%
Our definition of an \et\
matrix involves two properties which have implications for the
classical analogue of the quantum graph. The first ensures that a
classical particle on a directed bond has probability zero to be
back-scattered to the reversal of that bond in the next time step.
The second property produces democratic transmission probabilities
$M_{(pj)(mq)}=(v_j-1)^{-1}\delta_{mj}$, for $p\ne q$

Other common choices for vertex scattering matrices are the Neumann
matrix $\sigma^{[N]}$;
\begin{equation}
  \sigma^{[N]}_{pq} = \frac 2v-\delta_{pq},
\end{equation}
and Fourier transform matrices $\sigma^{[F]}$ introduced as vertex
scattering matrices in \cite{tan:usm},
\begin{equation} \label{eq:fourier}
\sigma^{[F]}_{pq} = \frac1{\sqrt{v}}\rme^{2\pi\rmi pq/v} \ .
\end{equation}
Fourier transform scattering matrices allow back-scattering with
equal transmission and reflection probabilities,
$M_{(pj)(mq)}=v_j^{-1}\delta_{mj}$. The Neumann scattering matrices
prefer back-scattering transitions over all other transitions 
put together.

The transition probabilities induced by the three classes of scattering
matrices at a vertex with a valency $v$, can  be summarised by
\begin{equation}\label{eq:bass:uno}
|\sigma_{ii}|^2 = r \qquad \mbox{and}\qquad |\sigma_{ij}|^2 =
\frac{1-r}{v-1} \quad\mbox{for $i\neq j$,}
\end{equation}
where $r=(2/v-1)^2$ (Neumann), $r=1/v$ (Fourier transform) and $r=0$
(\et).

The aims of the present manuscript can be summarised as follows:
\begin{enumerate}
\item
To explain the motivations for studying graphs with \et\ vertex
scattering matrices.
\item
To prove that \et\ vertex scattering matrices exist for arbitrarily
large $v$, and to provide practical methods to construct them.
\item
To investigate their spectral properties on both the classical (gap
estimates) and the quantum (spectral statistics) levels.
\end{enumerate}

The next section will explain the motivation for introducing the
\et\ scattering matrices. While \et\ matrices will be shown to have
desirable qualities, an elementary calculation shows it is trivial
to construct  $2\times 2$ examples, but  no $3\times 3$ \et\ matrix
exists. In section \ref{construction} we address the question of
their existence in other dimensions, and provide several infinite
sets of examples. Whether \et\ matrices exist in all dimensions
greater than three is an interesting open question.
 In section \ref{sec:properties} we compare spectra of the matrix $M$
for graphs quantised with different scattering matrices. We find
that in some fairly general situations the spectral gap in \et\
quantum graphs is larger than that obtained with other scattering
matrices. In section \ref{sec:numerics} we present the results of
numerical simulations using \et\ quantum graphs that show that
random matrix statistics are reproduced.

\section{Motivations}\label{sec:motivations}

In this section we shall try to explain the motivations for the
introduction of the \et\ vertex matrices.

A fundamental building block in the spectral theory for quantum
graphs and  a key tool in understanding spectral correlations
\cite{gnu:qga}, is the trace formula \cite{kot:pot,rot:lsl}. It
relates the quantum spectrum of a graph, $\{k_n\}_{n\in\Z}$,
to the length spectrum of its periodic orbits, expressed as an
identity of distributions:
\begin{equation} \label{eq:trace}
  \sum_{n\in\Z} \rme^{\rmi k_n u} = {\curlyL}\delta(u) +
\sum_{{\rm p.o.}} \frac{\ell_p}{r_p}\left(A_p\delta(u-\ell_p)+
\bar{A}_p\delta(u+\ell_p)\right).
\end{equation}
The first term on the right-hand side of \eqref{eq:trace} is the
Weyl term. $\curlyL$ denotes twice the total length of the graph,
and plays the r\^ole of the volume. The more interesting second term
is a sum which goes over classes of closed itineraries on the graph
equivalent up to cyclic permutations of the edges. We call these
periodic orbits of the graph. The metric length of an orbit is
denoted $\ell_p$, and $r_p$ is the number of times the orbit is a
repetition of a shorter one. The amplitude factor $A_p$ is the
product of all elements of the vertex scattering matrices
encountered as the orbit passes from one bond to the next.

Since \et\ matrices forbid back-scattering transitions, any orbit in
which a traversal of a bond is followed immediately by the traversal
of its reverse is eradicated from the sum. This significantly
reduces the number of orbits which need to be considered---in fact
the orbits that remain are exactly the closed geodesics considered
in combinatorial graph theory which will be discussed below. 
For a $v$-regular graph the asymptotic number of orbits of period
$n$ is reduced from $v^n/n$ to $(v-1)^n/n$.

We anticipate that using \et\ scattering matrices will significantly
simplify periodic orbit theories in quantum graphs, and perhaps lead
to new interesting problems and breakthroughs. The following will
serve as an example: A theorem of Gutkin and Smilansky \cite{gut:coh}
%
%
guarantees that one can ``hear'' the shape of a graph if the bond
lengths are rationally independent and if the vertex scattering
matrices are ``properly connecting''. The \et\ scattering matrices do
not belong to the latter class, and the question if one can ``hear''
\et\ graphs is open. The proof of the above mentioned theorem is
based on the trace formula, and relies heavily on the special
properties of the 2-periodic orbits which are absent from the trace
formula of \et\ graphs. Thus, a completely novel approach has to be
developed.

The study of quantum graphs with  \et\ vertex scattering matrices
leads to connections with objects that have been extensively studied
in combinatorial graph theory, such as the Ihara-Selberg zeta function
\cite{iha:ods}
and its generalisations. One way to see this
connection is by studying the spectrum of the classical evolution
operation $M$, which is defined as  the zero set of the secular
function
\begin{equation}
Z_M(\mu) \coloneq \det [I_{2B}-\mu M] \label{eq:clasec} \ .
 \end{equation}
This equation can be used to derive a classical (Rouelle like)
trace formula which is based on the periodic orbits on the graph
weighted by the products of scattering probabilities along the
orbit. When dealing with \et\ vertex scattering matrices, the trace
formula (and the corresponding zeta function) includes only periodic
orbits without back-scatter. These are the orbits which appear in
the Ihara-Selberg zeta function and its extensions \cite{sta:zff,sta:mzf, kot:zff,
hor:wzf}.
%
%
%
Moreover, the zeta functions are related by simple transformations,
so that an interesting correspondence between the two seemingly
unrelated problems can be established. We have mentioned previously
the classical spectral gap $\Delta$, and its conjectured influence
on the spectral statistics in the corresponding quantum graph.
Establishing bounds on $\Delta$ arises in the study of the Ihara-Selberg
zeta function, which makes an intriguing link between spectral
statistics and number theory. (See \cite{kea:rmL} for a review of the
connections between {\em arithmetical}\/ zeta functions and spectral
statistics.) We
shall make use of these connections here (see section \ref{sec:properties}).

 We are not able to prove that graphs with \et\
vertex scattering matrices display spectral statistics which
reproduce the predictions of random matrix theory. We present,
however, quite convincing numerical evidence showing that both their
spectral repulsion and  spectral rigidity adhere to the predictions
of the canonical random matrix ensembles. Whether this observation
is valid and can be rigorously formulated is an open problem
awaiting future research.

\section{Existence and construction } \label{construction}
In this section we shall show that the set of \et\ matrices is not
empty or trivial. It is easy to construct \et\ matrices in  dimension 2. It
is equally easy to show that \et\ matrices do not exist in dimension
3. We are not able to provide a list of the dimensions for which
\et\ matrices exist. We can, however, show that this list is
infinite. We do it by constructing examples of \et\ matrices using
skew-Hadamard matrices \cite{wil:hdt, joh:ist, whi:aif}, and
Dirichlet characters \cite{ire:aci}.

\subsection {Construction of \et\ matrices using Hadamard\\ matrices}

\begin{definition} \label{def:hadamard}
A Hadamard matrix $H$ is a matrix whose entries are $\pm 1$, and
whose columns are orthogonal. A skew-Hadamard matrix is a Hadamard
matrix satisfying the additional condition
$$ H+H^{\rm T} = 2 I_v,$$.
\end{definition}

\begin{proposition} \label{prop:exist}
Let $H$ be a  $v \times v$ skew-Hadamard matrix. Then
\begin{equation}
 \sigma = \frac{1}{\sqrt{v-1}}(H-I_v)
\end{equation}
is an \et\ matrix.
\end{proposition}
\dimostrazione All entries of $\sigma$ are $\pm(v-1)^{-1/2}$ except
for the zero entries along the diagonal. So we need only check
unitarity. But this is clear, since
\begin{align}
\sigma\sigma^{\rm T} &= \frac{1}{v-1}\left( HH^{\rm T}
- (H+H^{\rm T}) + I_v \right) \nonumber \\
&= I_v
\end{align}
since necessarily $HH^{\rm T}=v I_v$.\finire

Hadamard matrices have been conjectured to exist in dimensions $1,2$
and all multiples of $4$. This conjecture appears to date back to
\cite{pal:oom}. Various constructions of skew-Hadamard matrices are
known \cite{wil:hdt, joh:ist, whi:aif}. Currently there are
constructions of skew-Hadamard matrices for all dimensions which are
a multiple of 4 up to and including 184 \cite{dok:shm, geo:ogm},
plus other infinite sets of dimensions.

\subsection {Construction of \et\ matrices using Dirichlet\\ characters}

This method provides \et\ matrices of dimensions  $P+1$ where $P$ is prime.

\begin{proposition}
  \label{prop:character}
Let $P$ be an odd prime and let $\chi$ be a non-trivial Dirichlet
character modulo $P$. Let $C$ be the $P\times P$ matrix defined by
$C_{j\ell}=\chi(\ell-j)$. Then
\begin{equation}
\sigma =\frac1{\sqrt{P}}\left( \begin{array}{cc}
0 & \mbox{$1\;\cdots\;1$} \\
\begin{array}{c} 1 \\ \vdots \\1\end{array} & \mbox{\Huge$C$}
\end{array}\right)
\end{equation}
is \et.
\end{proposition}
We first need an auxiliary lemma.
\begin{lemma}
  Let $P$ be an odd prime, $\chi$ a non-trivial Dirichlet character,
and $j\in \Z$. Then
\begin{equation}
  \sum_{n=0}^{P-1} \chi(n-j)\overline{\chi(n)} = \left\{
\begin{array}{rl}
P-1 & \mbox{if $j\equiv 0\mod P$,} \\
-1 & \mbox{otherwise.}\end{array}\right.
\end{equation}
\end{lemma}
\dimostrazione If $j\equiv 0$ modulo $P$ then
\begin{equation}
  \sum_{n=0}^{P-1} |\chi(n)|^2 = \sum_{n=1}^{P-1} 1 = P-1.
\end{equation}
Otherwise, writing $n^{-1}$ for the multiplicative inverse of $n$ in
the finite field $\Z/P\Z$,
\begin{align}
 \sum_{n=0}^{P-1} \chi(n-j)\overline{\chi(n)} &= \sum_{n=1}^{P-1}
 \chi(n-j)\chi(n)^{-1} \nonumber \\
&=\sum_{n=1}^{P-1}\chi((n-j)n^{-1}) \nonumber \\
&=\sum_{n=1}^{P-1}\chi(1-jn^{-1}) \nonumber \\
\end{align}
Now since $n^{-1}$ runs over all invertible elements in the field,
the only argument which does not appear in the sum is $1$. So
\begin{align}
 \sum_{n=0}^{P-1} \chi(n-j)\overline{\chi(n)}
&=\sum_{m=0}^{P-1}\chi(m) - \chi(1) \nonumber \\
&=-1,
\end{align}
Using orthogonality of Dirichlet characters \cite{ire:aci} and the
known value $\chi(1)=1$.\finire

\dimostrazionea{proposition \ref{prop:character}} We first observe
that $|\chi(n)|=1$ unless $n\equiv 0$ whence $\chi(0)=0$, so
$\sigma$ has the required form. It is easy to see that unitarity
will follow once we prove that
\begin{equation} \label{eq:C:matrix}
CC^\dag = \left( \begin{array}{cccc}
P-1 & -1 & \cdots & -1 \\
-1 & P-1 & \cdots & -1 \\
\vdots & \vdots & \ddots & \vdots \\
-1 & -1 & \cdots & P-1
\end{array}\right).
\end{equation}
For this, note that the inner product of the $j^{\rm th}$ and
$\ell^{\rm th}$ columns of $C$ can be written
\begin{align}
\sum_{m=1}^P C_{mj}\overline{C_{m\ell}} &= \sum_{m=1}^P \chi(j-m)
\overline{\chi(\ell-m)} \nonumber \\
&=\sum_{n=0}^{P-1} \chi(n+j-\ell)\overline{\chi(n)},
\label{eq:char:sum}
\end{align}
via a change of index of summation. Now \eqref{eq:C:matrix} follows
by using the lemma to evaluate the sum \eqref{eq:char:sum}. \finire


\begin{corollary} \label{cor:dirichlet}
Let $P$ be a prime congruent to $1$ modulo $4$. Then there exists a
symmetric \et\ matrix of dimension $P+1$.
\end{corollary}
\dimostrazione We use the construction in proposition
\ref{prop:character} with the Legendre symbol as the Dirichlet
character $\chi(n)=\left (\frac{n}{P}\right)$ where
\begin{equation}
  \left(\frac{n}{P}\right) = \left\{
\begin{array}{rl}
0 & \mbox{if $n\equiv 0\mod P$,} \\
1 & \mbox{if $n$ is a square modulo $P$,}\\
-1 & \mbox{if $n$ is not a square modulo $P$.}
\end{array}\right.
\end{equation}
To show that $\sigma$ is symmetric it suffices to show that the
circulant matrix $C$ is symmetric. To see this, note that
\begin{equation*}
  \chi(P-m) = \left(\frac{m(-1)}P\right) = \left(\frac{m}P\right)\left(\frac
{-1}P\right) = \left(\frac{m}P\right)=\chi(m)
\end{equation*}
since $(-1/P)=1$ if $P\equiv 1$ modulo $4$ (Euler).\finire

By Dirichlet's theorem, corollary \ref{cor:dirichlet} provides
infinitely many examples of symmetric \et\ matrices.

The constructions in propositions \ref{prop:exist} and
\ref{prop:character} give many examples of \et\ matrices of {\em
even} dimensions. We can construct an example of an \et\ matrix in
dimension $5$,
\begin{equation} \label{eq:five}
\sigma=\frac12\left(
\begin{array}{ccccc}
0 & 1 & 1 & 1 & 1 \\
1 & 0 & 1 & \omega & \omega^2 \\
1 & 1 & 0 & \omega^2 & \omega \\
1 & \omega & \omega^2 & 0 & 1 \\
1 & \omega^2 & \omega & 1 & 0
\end{array}\right), \qquad\mbox{where $\omega=\rme^{2\pi\rmi/3}$.}
\end{equation}
However, apart from this example  we do not have any examples in
{\em odd}\/ dimensions. These \et\ matrices do not appear to have
been studied in the literature. An examination of \eqref{eq:five}
might lead one to suspect that such matrices can be constructed for
dimension $v$ by using entries that are $(v-2)^{\rm th}$ roots of
unity. However we have exhaustively checked for the case $v=7$ and
shown this to be false.

\section{Properties of graphs with \et\ scattering\\ matrices}
\label{sec:properties}

Now that we have demonstrated the existence of \et\ matrices, we can study
the quantum and classical evolutions which they induce. In
particular, we shall use some results from
combinatorial graph theory to demonstrate the advantages gained by
studying graphs with \et\ vertex scattering matrices.

An important tool in the preceding discussion is the spectrum of the
connectivity matrix $C$ defined in the introduction.  Since $C$ is
symmetric its eigenvalues $\mu_j$ are real, and we order them:
$\mu_{V-1}\leq\cdots \leq\mu_1\leq\mu_0$.

We shall consider  $v$-regular graphs and  to avoid trivial cases we
will assume throughout that $v> 3$. A connected $v$-regular graph
has the property that its connectivity matrix has largest eigenvalue
$\mu_0=v$ and it is simple. A $v$-regular graph is called {\em
Ramanujan} if all other eigenvalues of $C$ are contained in the
interval $[-2\sqrt{v-1},2\sqrt{v-1}]$. Ramanujan graphs are of
interest in computer science and communication network theory since
they are sparse yet highly connected \cite{dav:ent,mur:rg}.

We shall discuss the spectrum of the classical evolution operator
$M$ on a  $v$-regular graph constructed by using vertex scattering
matrices of the types listed in (\ref{eq:bass:uno}). We shall relate
them to the spectrum of the connectivity matrix $C$ by the following
theorem.

\begin{theorem} \label{thm:uno}
Let $M$ be the doubly stochastic transition probabilities matrix
associated to a quantum $v$-regular graph with unitary vertex
scattering matrices $\sigma$ which satisfy \eqref{eq:bass:uno} for
some $r>0$. Let the eigenvalues of $C$ be $\mu_0\geq \mu_1\geq
\cdots \geq \mu_{V-1}$. Let
\begin{equation*}
  u_j\coloneq\frac{(1-r)\mu_j +\sqrt{(1-r)^2\mu_j^2-4(1-rv)(v-1)}}{2(v-1)}
\end{equation*}
and
\begin{equation*}
  \tilde{u}_j\coloneq
\frac{(1-r)\mu_j -\sqrt{(1-r)^2\mu_j^2-4(1-rv)(v-1)}}{2(v-1)}.
\end{equation*}
The spectrum of $M$ consists of the points
\begin{equation*}
  u_0,\ldots,u_{V-1},\tilde{u}_0,\ldots,\tilde{u}_{V-1},\frac{|1-rv|}{v-1},
-\frac{|1-rv|}{v-1}
\end{equation*}
where the last two points are listed with multiplicity
$\displaystyle \frac{(v-2)V}2$.
\end{theorem}

\dimostrazione Let $W$ be the matrix representing Hashimoto's bond
(edge) adjacency operator \cite{has:zff,hor:wzf} defined as:
\begin{equation}
W_{(pj)(mq)} = \delta_{mj} (1-\delta_{pq}).
\end{equation}
It has entries equal to $1$ only when two directed bonds 
follow each other at a common vertex $j$, but excluding back-scattering.
The form of the vertex scattering matrices $\sigma$ of interest here
is provided by (\ref{eq:bass:uno}). It implies that for $v$-regular
graphs,
\begin{equation*}
  M = \left(\frac{1-r}{v-1}\right)W  + r J,
\end{equation*}
where $r=(2/v-1)^2$ (Neumann), $r=1/v$ (Fourier transform) and $r=0$
(\et).  The characteristic polynomial of $M$, $\det[uI_{2B}-M]$ is
related to a graph theoretic zeta function developed in
\cite{bar:cpg}. Bartholi's theorem \cite{bar:cpg,miz:anp} implies an
equivalent form for the characteristic polynomial in terms of the
matrix $C$,
\begin{equation*}
  \det[uI_{2B}-M] = \left( u^2-\frac{(1-rv)^2}{(v-1)^2}\right)^{(v-2)V/2}
\det\left[\left(u^2+\frac{1-rv}{v-1}\right)I_V
-\frac{1-r}{v-1}Cu\right].
\end{equation*}
It follows that eigenvalues of $M$ are solutions to
\begin{equation*}
  u^2-\frac{(1-rv)^2}{(v-1)^2} = 0,
\end{equation*}
(with multiplicity $(v-2)V/2$) and
\begin{equation*}
  u^2-\frac{1-r}{v-1}\mu_j u +\frac{1-rv}{v-1} = 0,
\end{equation*}
for $j=0,\ldots,V-1$.\finire

For regular graphs with \et\ scattering matrices, theorem \ref{thm:uno}
shows that the eigenvalues of $M$ are (up to a scaling) at the positions
of the poles of the Ihara-Selberg zeta function \cite{iha:ods} of the
graph, as was noted in \cite{smi:qcd}. The proof in this case follows from
Bass' identity \cite{bas:tis} of which Bartholi's theorem
is a generalisation.

Theorem \ref{thm:uno} will enable us to compare \et\ scattering matrices
with others of type \eqref{eq:bass:uno} (see Theorem \ref{thm:due} below).
It also has a number of other consequences which may be of independent
interest.

\begin{corollary} \label{cor:ramanujan}
\begin{itemize}
\item If $r\geq 1/v$ then all eigenvalues of $M$ are real.
\item If $r=1/v$  (e.g.\ Fourier transform scattering matrix) then the 
eigenvalues of $M$ are
$$\frac{\mu_0}v,\ldots,\frac{\mu_{V-1}}v,0$$
and the $0$ has multiplicity $(v-1)V$.
\item If $r < 1/v$ then $u_j$ and $\tilde{u}_j$ are real iff $|\mu_j|\geq
\displaystyle \frac2{1-r}\sqrt{(1-rv)(v-1)}$.
\item If $r <1/v$ and $|\mu_j|<\displaystyle \frac2{1-r}\sqrt{(1-rv)(v-1)}$ 
then $$|u_j|=|\tilde{u}_j| = \sqrt{\frac{1-rv}{v-1}}.$$
\end{itemize}
\end{corollary}
In particular, for a $v$-regular graph which is Ramanujan, and has
\et\ scattering matrices, all but the eigenvalue $1$ lie in a disc
of radius $(v-1)^{-1/2}$ about the origin.



\begin{theorem} \label{thm:due}
Consider a $v$-regular graph. Let the spectral gap for the quantum
graph with \et\ scattering matrices be $\Delta_{\rm et}$ and denote
by $\Delta_r$ the spectral gap for the same graph with scattering
matrices $\sigma$ satisfying \eqref{eq:bass:uno} for some $r>0$.
Then there exists $\epsilon>0$ such that if $C$ has an eigenvalue in
either of the intervals $(2\sqrt{v-1}-\epsilon,v)$ or
$[-v,-2\sqrt{v-1}+\epsilon)$ then $\Delta_{\rm et} > \Delta_r$.
\end{theorem}
We give the proof of theorem \ref{thm:due} at the end of this
section. Theorem \ref{thm:due} demonstrates that in some fairly
general situations the spectral gap arising in quantum graphs with
\et\ scattering matrices is larger than the spectral gap with other
kinds of scattering matrices. Notice that if one considers a
sequence of $v$-regular graphs with $V\to\infty$ the Alon-Boppana
bound \cite[Theorem 1.3.1]{dav:ent} states that
$$
\liminf_{V\to\infty} \mu_1 \geq 2\sqrt{v-1}.
$$
In other words, theorem \ref{thm:due} will apply eventually.

In the interests of full disclosure we point out that for
some other families of graphs \et\ matrices will not necessarily lead
to a larger (although still large) spectral gap, e.g.\ fully-connected
graphs.

Before giving the proof, we provide one more lemma.
\begin{lemma} \label{lem:increasing}
  Let $0\leq\mu < v$. Then define
$$f_\mu(r)\coloneq \frac{(1-r)\mu + \sqrt{(1-r)^2\mu^2 +
4(vr-1)(v-1)}}{2(v-1)}.$$ Then $f_\mu(r)$ is real and strictly
increasing on $\displaystyle 1-\frac{2(v-1)}{\mu^2}
(v-\sqrt{v^2-\mu^2}) \leq r \leq 1$.
\end{lemma}
\dimostrazione It is convenient to define a new variable
$X=v(1-r)/(v-1)$, so that $f_\mu$ becomes, after some manipulation,
$$
f_\mu(X)=\frac1{2v}\left( X\mu +
\sqrt{\mu^2X^2+4v^2(1-X)}\right),$$ and is real if $0\leq
X\leq\frac{2v^2-2v\sqrt{v^2-\mu^2}}{\mu^2}$. Differentiating,
\begin{align*}
  \frac{\rmd f_\mu}{\rmd X} &= \frac1{2v}\left( \mu - \frac{2v^2-\mu^2X}{\sqrt{
\mu^2X^2+4v^2(1-X)}}\right) \\
&\leq \frac1{2v}\left( \mu-\frac{(2-X)v^2}{\sqrt{v^2(X-2)^2}}\right) \\
&=\frac1{2v}(\mu-v) < 0
\end{align*}
since $\sqrt{(X-2)^2}=2-X$, as $0\leq X < 2$. As $\displaystyle
\frac{\rmd X}{\rmd r}<0$ it follows that $f_\mu(r)$ is strictly increasing.
\finire

\dimostrazionea{theorem \ref{thm:due}}
We assume $|\mu_1|\geq|\mu_{V-1}|$. If this is not the case then the
argument below holds {\it mutatis mutandis}\/ replacing $\mu_1$ by
$\mu_{V-1}$.

If the graph is Ramanujan, and if $r<1/v$, then choose $\epsilon$ so
that
\begin{equation}
  \frac{2-rv}{1-r} \sqrt{v-1}
< 2\sqrt{v-1}-\epsilon.
\end{equation}
Since $v>3$ if $r>0$ such an $\epsilon$ can always be found.
 
As $\displaystyle \mu_1> \frac{2-rv}{1-r} \sqrt{v-1} > \frac2{1-r}\sqrt{(1-rv)(v-1)}$ by corollary 
\ref{cor:ramanujan} $u_1=f_{\mu_1}(r)$ is real and $\Delta_r=1-u_1$, where
\begin{equation}
  u_1 = \frac{(1-r)\mu_1 +\sqrt{(1-r)^2\mu_1^2 - 4(1-rv)(v-1)}}{2(v-1)}
  > \frac{1}{\sqrt{v-1}}.
\end{equation}
For Ramanujan graphs, $\displaystyle \Delta_{\rm
et}=1-\frac1{\sqrt{v-1}}
>\Delta_r$.

If the graph is Ramanujan and $r\geq1/v$ then choose $\epsilon$ to
satisfy
\begin{equation}
  \frac{v}{\sqrt{v-1}}<2\sqrt{v-1}-\epsilon.
\end{equation}
By lemma \ref{lem:increasing},
\begin{equation}
  u_1 \geq  f_{\mu_1}(1/v) = \frac{\mu_1}v > \frac1{\sqrt{v-1}},
\end{equation}
and again we have $\Delta_{\rm et}>\Delta_r$.

If the graph is not Ramanujan, then
$u_1$ is still real since
$$\mu_1>2\sqrt{v-1}>\displaystyle
\frac{2}{r-1}\sqrt{(1-rv)(v-1)}\qquad\mbox{if $r<1/v$,}$$ so by
lemma \ref{lem:increasing}
\begin{equation}
1-\Delta_r =f_{\mu_1}(r) 
>f_{\mu_1}(0) 
=1-\Delta_{\rm et}.
\end{equation}
\finire




\section{Numerical simulations}\label{sec:numerics}
The results of the previous section show that well-connected quantum
graphs with \et\ scattering matrices can have large spectral gaps.
According to the Tanner conjecture \cite{tan:usm} the spectral
statistics should converge to the statistics of ensembles of random
matrix theory. To illustrate this we present the results of some
numerical calculations of the nearest-neighbour spacing density
$P(s)$ for points in the spectrum of a quantum graph, and the
variance $V(L)$ of the number of points in an interval of length
$L$.

To calculate these spectral statistics we did not solve
\eqref{eq:spec:det} directly. Rather we took an approach which is
known to be equivalent. We replaced the phases $kL_{(pj)}$ in
\eqref{eq:defnU} by random phases in the interval $[0,2\pi]$ and
diagonalised the resulting matrices. The statistics of the re-scaled
eigen-phases approach those of the spectrum defined by
\eqref{eq:spec:det} in the limit as $B\to\infty$ if the bond lengths
are not rationally related, and drawn from a narrowing interval as
$B\to\infty$ \cite{gnu:qga,ber:xxx}.

\begin{figure}[h]
\begin{center}
\setlength{\unitlength}{5cm}
\begin{picture}(2.3,1)
 \put(0,0){\includegraphics[angle=0,width=5cm,height=5cm]{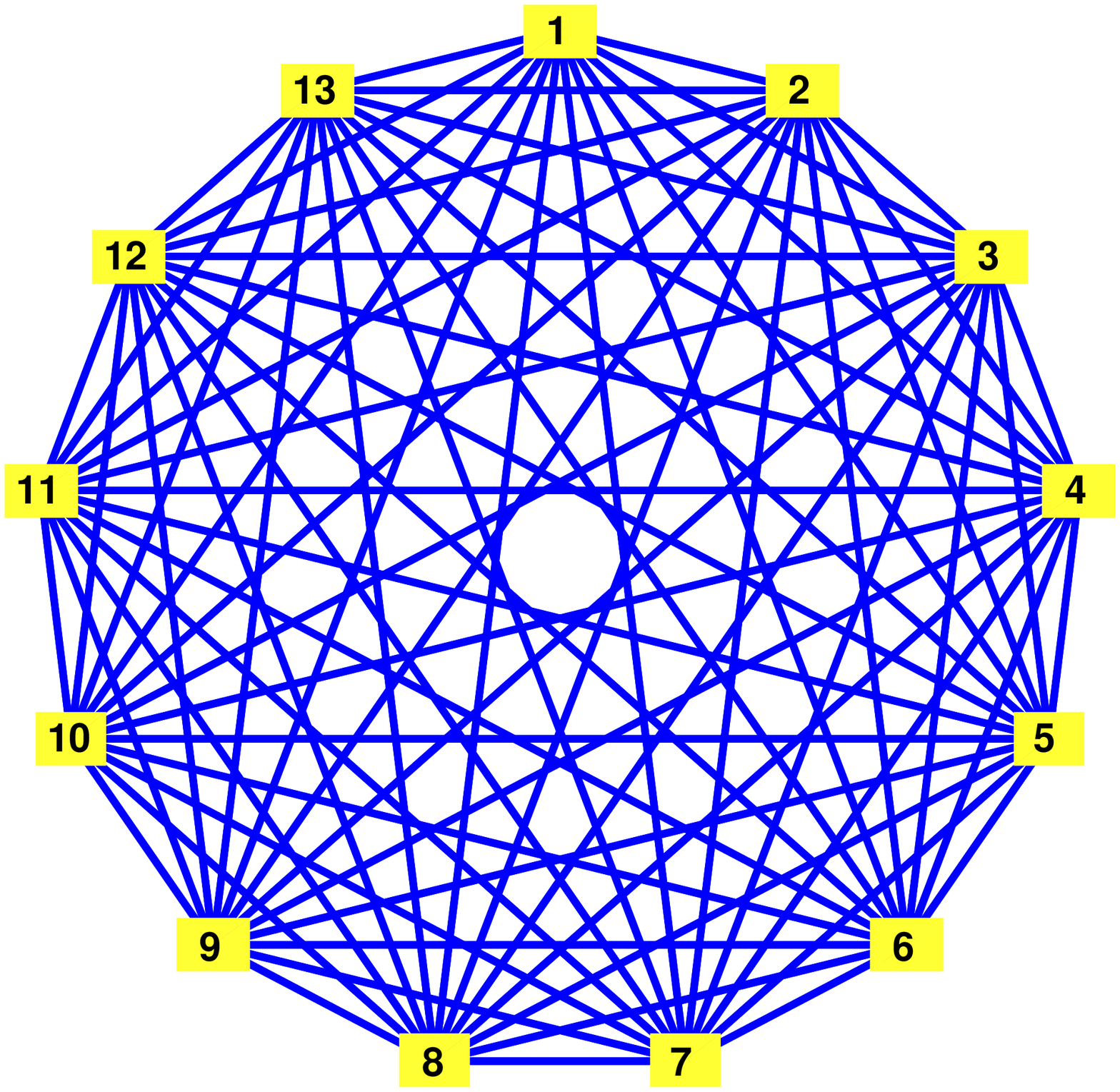}}
 \put(1.3,0){\includegraphics[angle=0,width=5cm,height=5cm]{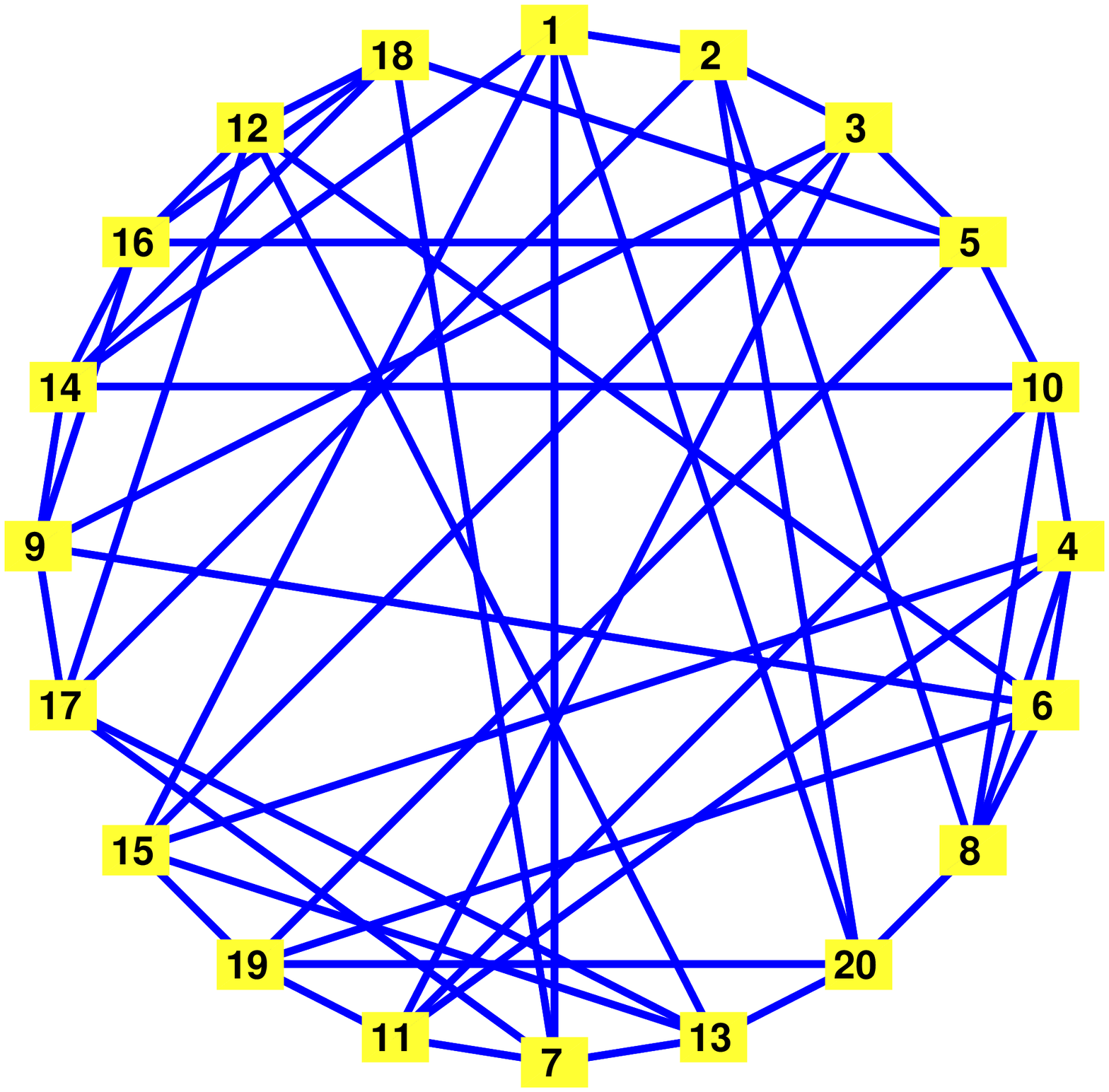}}
\put(0,0.9){a)} \put(1.3,0.9){b)}
\end{picture}

\caption{The graphs used in the numerical simulations: a) A complete
graph with 13 vertices, b) a $5$-regular graph with 20 vertices.}
\label{fig:uno}
\end{center}
\end{figure}

\begin{figure}[h]
\begin{center}
\setlength{\unitlength}{5cm}
\begin{picture}(3,1.3)
\put(0,0.1){\includegraphics[angle=0,width=7.0cm,height=6cm]{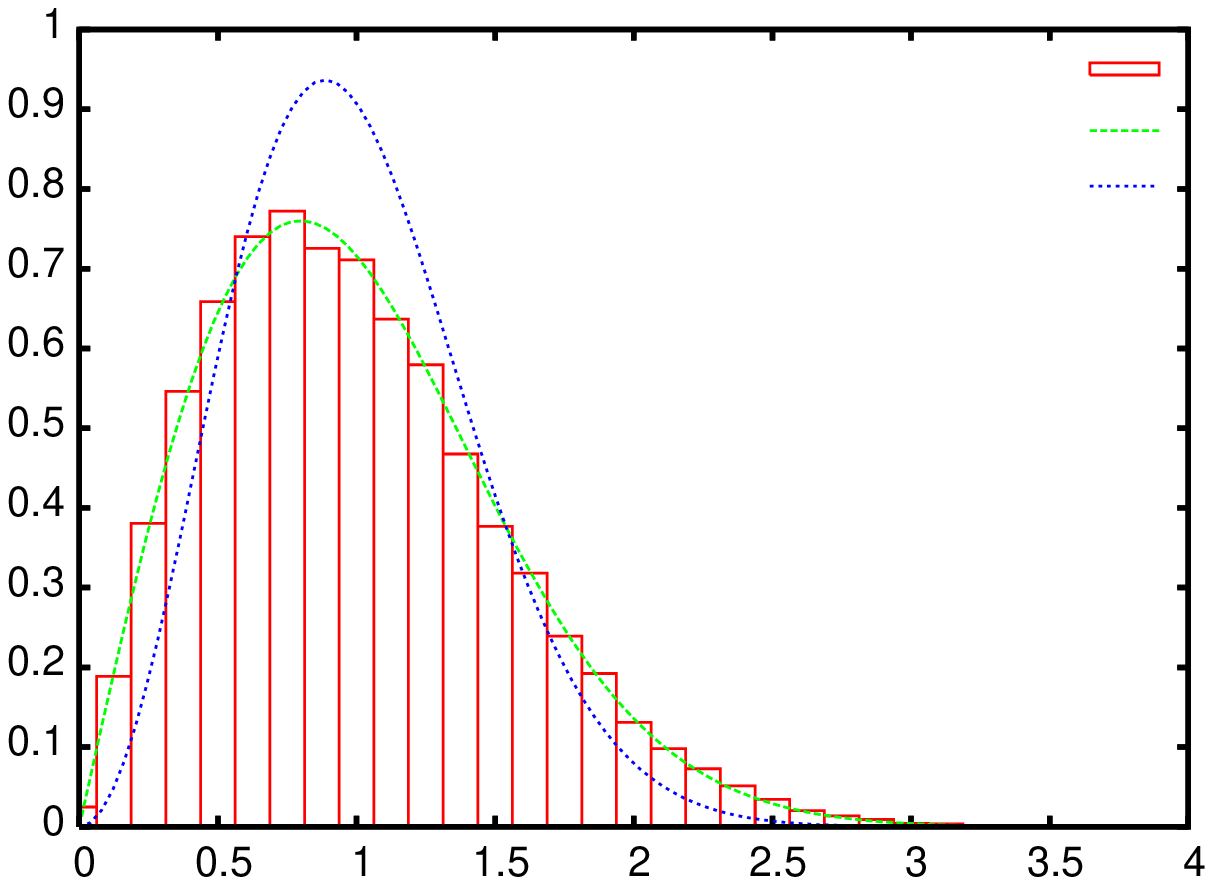}}
\put(1.5,0.1){\includegraphics[angle=0,width=7.0cm,height=6cm]{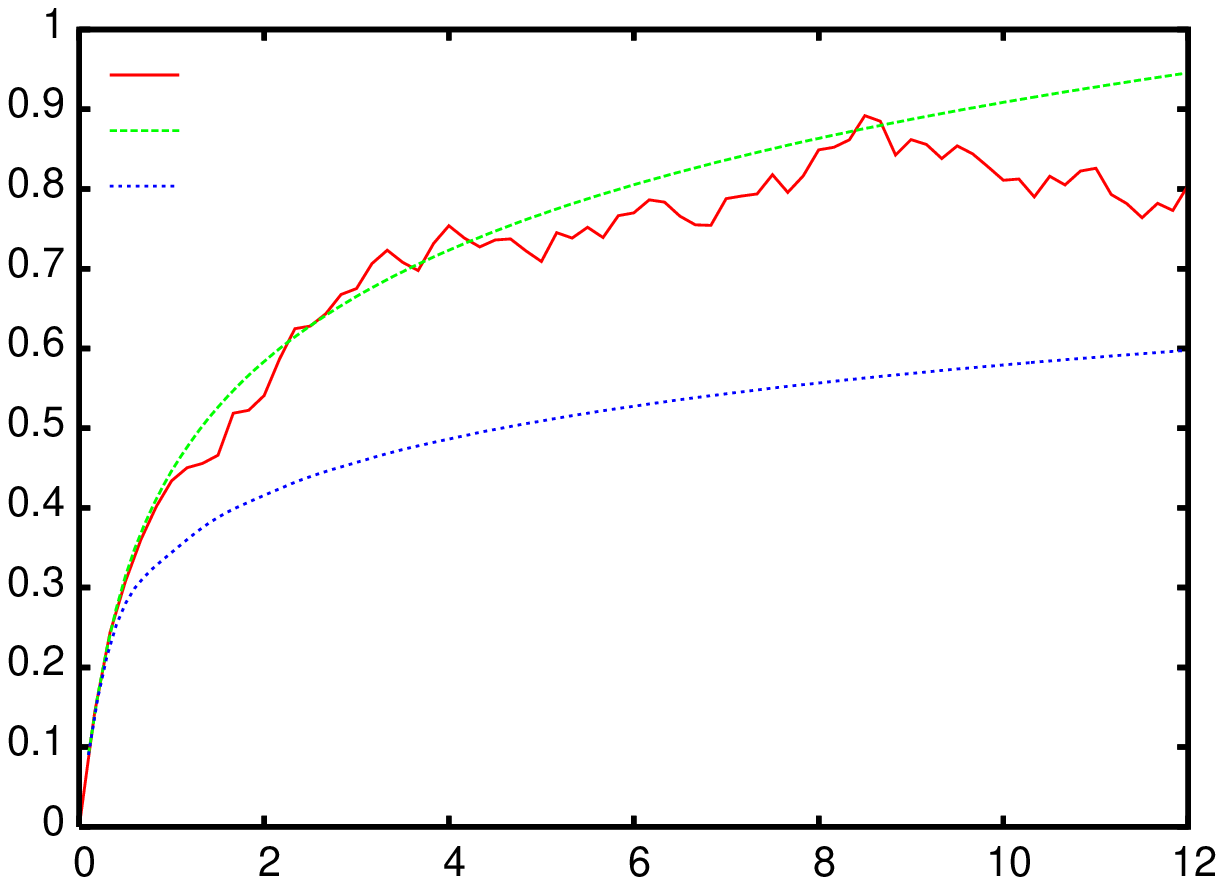}}
\put(0.721,1.19){{\small Quantum graph}} \put(1.06,1.10){{\small
\E{}OE}} \put(1.06,1.01){{\small \E{}UE}} \put(1.75,1.19){{\small
Quantum graph}} \put(1.75,1.10){{\small \E{}OE}}
\put(1.75,1.01){{\small \E{}UE}} \put(-0.1,1.21){$P(s)$}
\put(1.3,0.04){$s$} \put(2.8,0.04){$L$} \put(1.4,1.21){$V(L)$}
\end{picture}
\caption{Spectral statistics for a quantum regular graph with symmetric scattering matrices. On the left is a plot of the
nearest-neighbour density; on the right is the number variance.}
\label{fig:due}
\end{center}
\end{figure}

In figure \ref{fig:due} we plot the nearest neighbour distribution
and number variance for the $5$-regular graph in figure
\ref{fig:uno}b) with the \et\ vertex scattering matrix given
explicitly in (\ref{eq:five}). Since $\sigma$ is symmetric we expect
the spectral statistics to approach those of the \ensemble\
orthogonal ensemble (\E{}OE) of random matrices. We also plot in
figure \ref{fig:due} the corresponding limiting curves for
\ensemble\ orthogonal ensemble and \ensemble\ unitary ensemble
(\E{}UE) \cite{meh:rm}. We see, as expected, agreement to the
\ensemble\ orthogonal ensemble curves, even for such a relatively
small graph.

\begin{figure}[h]
\begin{center}
\setlength{\unitlength}{5cm}
\begin{picture}(3,1.3)
\put(0,0.1){\includegraphics[angle=0,width=7.0cm,height=6cm]{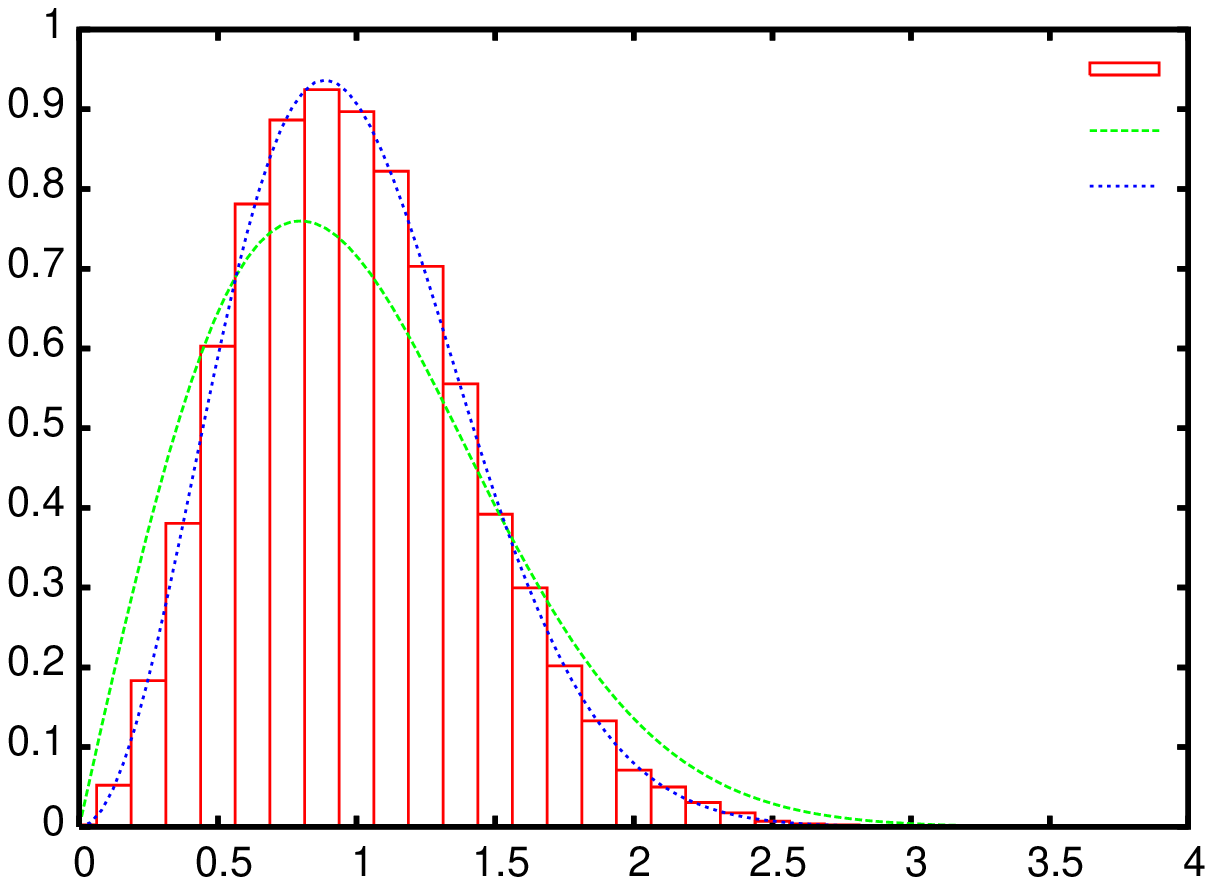}}
\put(1.5,0.1){\includegraphics[angle=0,width=7.0cm,height=6cm]{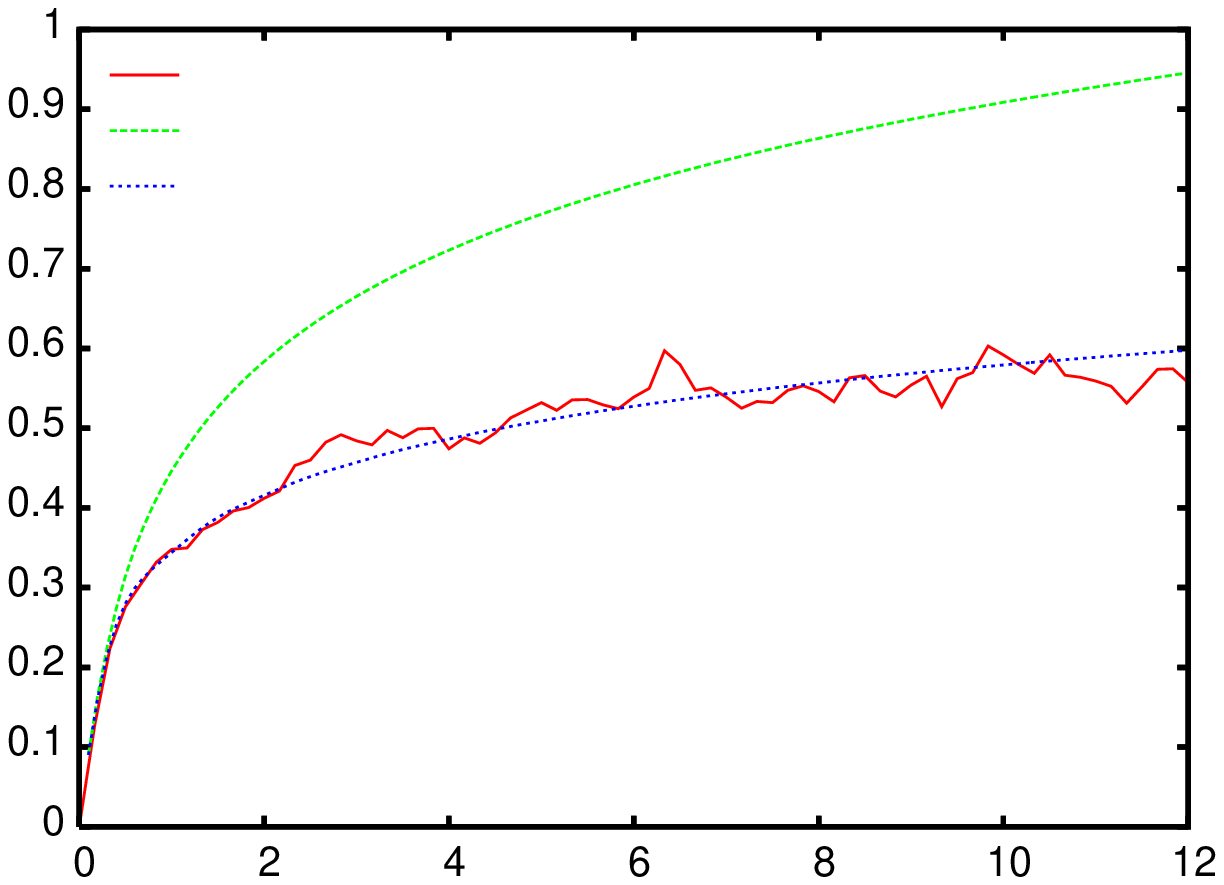}}
\put(0.721,1.19){{\small Quantum graph}} \put(1.06,1.10){{\small
\E{}OE}} \put(1.06,1.01){{\small \E{}UE}} \put(1.75,1.19){{\small
Quantum graph}} \put(1.75,1.10){{\small \E{}OE}}
\put(1.75,1.01){{\small \E{}UE}} \put(-0.1,1.21){$P(s)$}
\put(1.3,0.04){$s$} \put(2.8,0.04){$L$} \put(1.4,1.21){$V(L)$}
\end{picture}
\caption{Spectral statistics for the quantum complete graph with anti-symmetric scattering matrices. On the left is a plot of
the nearest-neighbour density; on the right is the number variance.}
\label{fig:tre}
\end{center}
\end{figure}

In figure \ref{fig:tre} we present the corresponding numerics for the
complete graph on 13 vertices (figure 1a) 
with an anti-symmetric \et\ scattering matrix
at each vertex. In this case we expect convergence to the \ensemble\ unitary
ensemble statistics, and this is also clearly demonstrated in the figure.

Given the utility of these \et\ matrices, we believe the study of their
existence in dimensions for which we do not currently have examples
is interesting and merits further investigation.

\subsection*{Acknowledgements}
We are grateful for interesting discussions with
Gregory Berkolaiko and Nicole Raulf.

The work of JMH and BW is supported by the National
Sciences Foundation under research grant DMS-0604859. US
acknowledges support from the Minerva Center for non-linear Physics,
the Einstein (Minerva) Center at the Weizmann Institute and BSF
grant 2006065.

Part of this work was carried out while the authors were visiting
the Isaac Newton Institute for Mathematical Sciences, Cambridge, UK.
US acknowledge the EPSRC grant 531174 which supported his stay.
The stay of JMH and BW was partially supported by National Sciences
Foundation grant DMS-0648786.

\def\Dbar{\leavevmode\lower.6ex\hbox to 0pt{\hskip-.23ex \accent"16\hss}D}

\end{document}